\documentclass[pra,aps,twocolumn]{revtex4}
\usepackage{graphicx}
\usepackage{times}
\usepackage{soul}
\usepackage{color}
\usepackage{bm}
\usepackage{amsmath, amsthm, amssymb,mathrsfs}
\usepackage{hyperref}
\usepackage{float}

\input epsf

\def\pmb#1{\setbox0=\hbox{#1}%
   \kern-0.015em\copy0\kern-\wd0
   \kern-0.025em\copy0\kern-\wd0
   \kern-0.015em\raise.0233em\box0}

\begin{document}
\title[Spin entanglement in elastic e-Li collisions]
{Spin entanglement in elastic electron scattering from lithium atoms}

\author{
K.~Bartschat\footnote{Corresponding Author: klaus.bartschat@drake.edu} and S.~Fonseca dos Santos
}

\affiliation{Department of Physics and Astronomy, Drake University, Des Moines, Iowa 50311, USA}

\setstcolor{red}
\date{\today}


\begin{abstract}
In two recent papers (Phys.\ Rev.\ Lett.~{\bf 116} (2016) 033201; Phys.\ Rev.\ A~{\bf 94} (2016)  032331), 
the possibility of continuously varying the degree of entanglement between an elastically scattered electron and
the valence electron of an alkali target was discussed.
In order to estimate how well such a scheme may work in practice, we present 
results for elastic electron scattering from lithium in the energy regime of 1$-$5~eV and the full range
of scattering angles~$0^\circ - 180^\circ$.  The most promising regime for Bell-correlations in this particular
collision system are energies between about 1.5~eV and 3.0~eV, in an angular range around $110^\circ \pm 10^\circ$.
In addition to the relative exchange asymmetry parameter, we present the differential cross section that 
is important when estimating the count rate and hence the feasibility of experiments using this system. 
\end{abstract}

\maketitle


In two recent publications, Blum and Lohmann~\cite{BL2016} and Lohmann {\it et al.}~\cite{LBL2016} discussed a tunable entanglement
in elastic electron collisions with atomic hydrogen or light alkali atoms, where explicitly spin-dependent interactions
may be neglected and the process is completely described by two independent parameters, namely the absolute angle-differential
cross section (DCS) and a spin-correlation parameter~$P$, which (except for the opposite sign) is
the exchange asymmetry~$A_{\rm ex}$ that was measured by the Bielefeld~\cite{Baum-AexLi,Baum-Aex} and NIST~\cite{NIST-Aex} groups many years
ago.  Due to the available experimental data, Lohmann {\it et al.}~\cite{LBL2016} presented results for $P = -A_{\rm ex}$ 
from various close-coupling calculations for atomic hydrogen and sodium, but only for selected energies as a function of 
the scattering angle, and for lithium, but only at selected angles as a function of energy.  

In order to get a firm estimate whether such collision systems might be appropriate and also what the expected signal rate might be,
it is highly desirable to have a comprehensive dataset on an energy-angle grid.  Furthermore, a critical parameter
for practical applications is the absolute angle-differential cross section, since it determines whether or not the signal
rate is sufficient in an actual experiment.

\smallskip 

The results reported below were obtained in a \hbox{5-state} close-coupling model for \hbox{e-Li} collisions,
including the $(1s^2 2s)^2S$, $(1s^2 2p)^2P^{\rm o}$, $(1s^2 3s)^2S$, $(1s^2 3p)^2P^{\rm o}$, and
$(1s^2 3d)^2D$ states of Li in the close-coupling expansion.
For the collision energies of interest, such a simple model is sufficient, as we will demonstrate
by comparing its predictions with the few experimental data that are currently available.
The close-coupling equations were solved with the Belfast \hbox{R-matrix} code~\cite{BEN1995}, which has the
advantage of being able to handle many collision energies very efficiently, once the inner-region problem has
been solved by diagonalizing the hamiltonian matrices.  Specifically, we set the R-matrix radius
to 40~$a_0$, where $a_0 = 0.529 \time 10^{-10}\,$m is the Bohr radius.  
We calculated partial waves up to a total orbital angular momentum of $L=15$ and employed 25 continuum
orbitals to expand the \hbox{R-matrix} basis functions.

The essential point regarding spin entanglement is the following~\cite{BL2016,LBL2016}:  After the collision, the projectile and the 
target valence electron are correlated.  Depending on the collision energy and the scattering angle, the correlation can be 
classified by the value of~\cite{Baum-AexLi}
\begin{equation}
P = -A_{\rm ex} = \frac{\sigma^{\uparrow\uparrow} - \sigma^{\uparrow\downarrow}}{\sigma^{\uparrow\uparrow} + \sigma^{\uparrow\downarrow}} = 
                   \frac{\sigma^t-\sigma^s}{\sigma^t+\sigma^s}
\end{equation} 
where $\sigma^{\uparrow\uparrow}\,(\sigma^{\uparrow\downarrow})$ and $\sigma^s\,(\sigma^t)$ 
are short-cut notations (instead of $d\sigma(E,\theta)/d\Omega$) for the angle-differential cross sections (DCSs) for parallel (anti-parallel) 
spin orientations of the projectile and target spins or triplet (singlet) scattering.
The DCS for unpolarized projectile and target beams is given by 
\begin{equation}
\sigma_u = \frac{1}{4}\sigma^s + \frac{3}{4}\sigma^t,
\end{equation}
and hence the limiting values for $P$ are $+1/3$ for pure triplet and $-1$ for pure singlet scattering.
The latter extreme case corresponds to the well-known situation of two spins forming a 
combined spin-0 system.

According to Blum and Lohmann, the combined projectile + target spin system can be classified as 
separable~(S) if $P > -1/3$, entangled~(E) for $-1/3 > P > -1/\sqrt{2}$, or Bell-correlated (B) if $P < -1/\sqrt{2}$.
Hence, such systems may provide a knob to ``dial in'' the amount of correlation one would like.  Note that
it is not necessary for Bell correlations to have a pure singlet state.
  
\smallskip   

\begin{figure}
\begin{center}
\includegraphics[width=0.97\columnwidth]{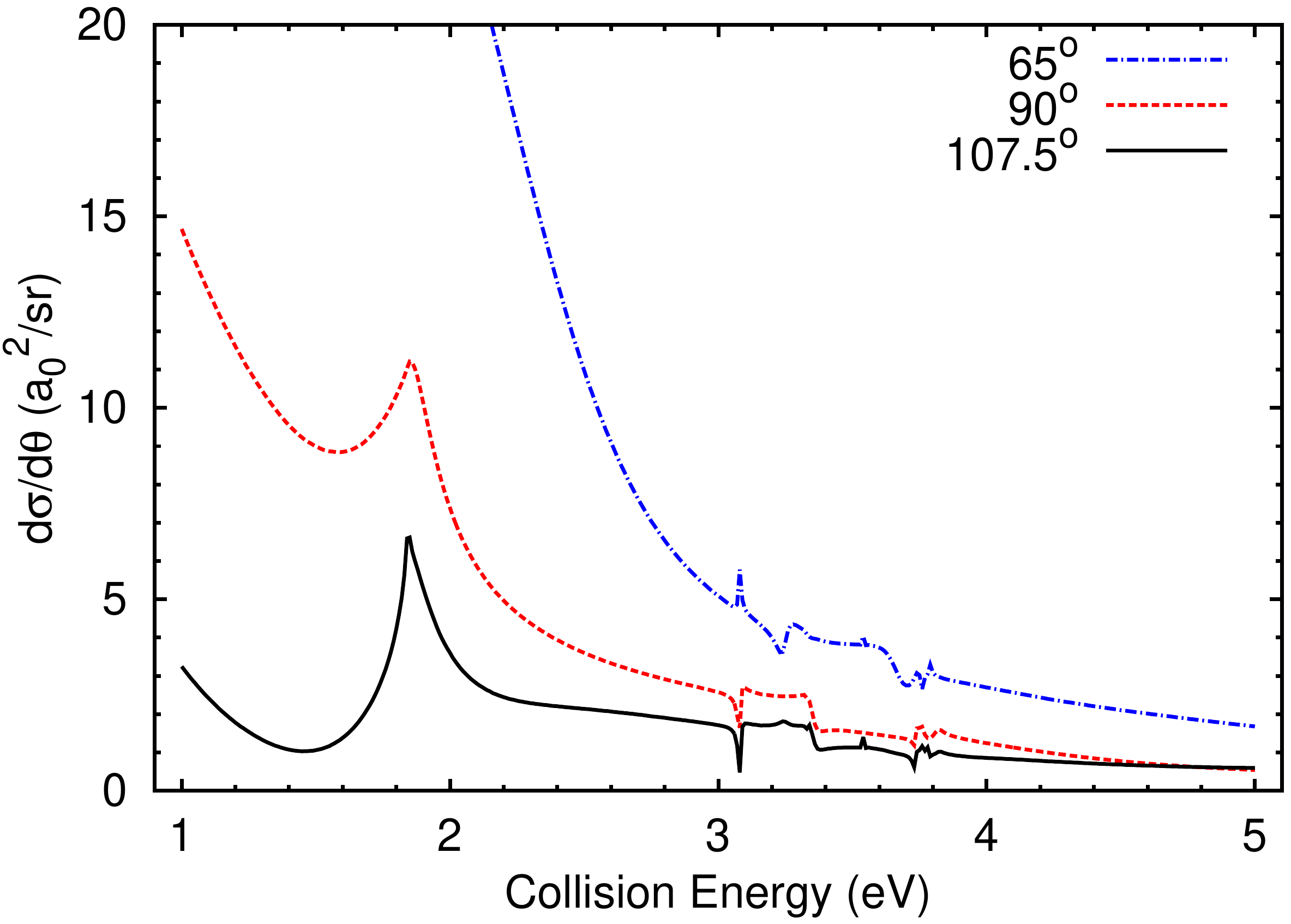}\\
~~~\includegraphics[width=0.95\columnwidth]{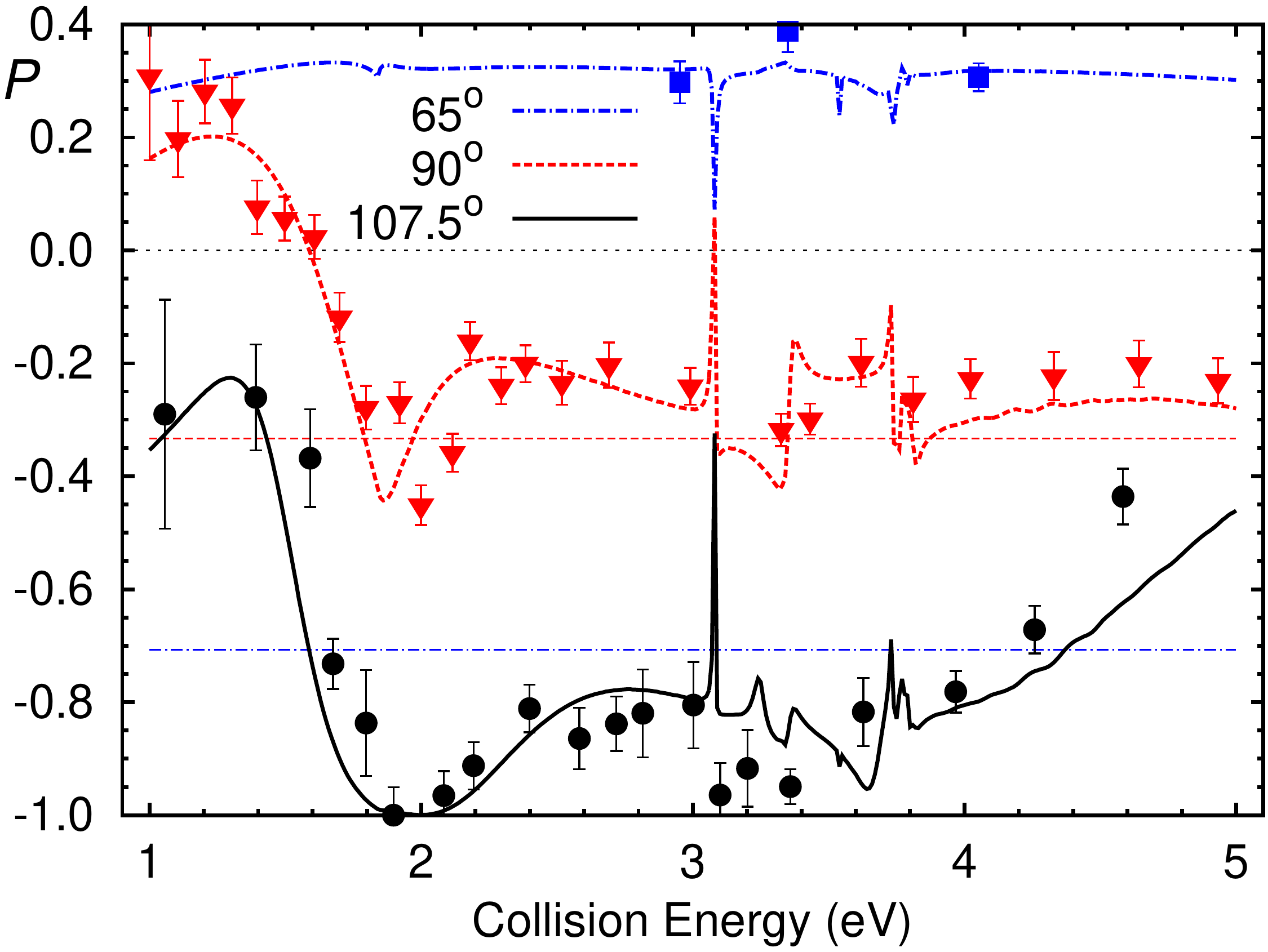}
\end{center}
\caption{Differential cross section (top) and spin correlation parameter~$P$ (bottom)
for elastic electron scattering from Li atoms as a function of energy at
scattering angles of $65^\circ$, $90^\circ$, and $107.5^\circ$.
The lines at $-1/3$ and $-1/\sqrt2$ in the panel for~$P$ mark the
borders between separable and entangled as well as entangled and Bell-correlated regions, respectively.
The experimental data for $P = -A_{\rm ex}$ are taken from Baum {\it et al.}~\cite{Baum-AexLi}.
}
\label{fig:energy}
\end{figure}

Figure~\ref{fig:energy} shows that the present 5-state model is, indeed, sufficient for the problem at hand.  The overall agreement 
with the available experimental in the energy range of $1-5$~eV is very satisfactory.  Since Bell correlations appear to be only 
realizable in this energy regime, it is not necessary to include coupling to higher Rydberg states or even the ionization continuum.
Note, however, the appearance of resonances in both the DCS and, even more pronounced, in~$P$.  Although these resonances would likely
be washed out in practice due to the finite energy resolution (they are real, but not visible in Fig.~7 of~\cite{LBL2016}, presumably due to 
the energy grid chosen in the calculations), it seems advisable to avoid the resonance regime from 3~eV up to the ionization threshold
when choosing the energy.

\begin{figure}[H]
\bigskip
\includegraphics[width=0.99\columnwidth]{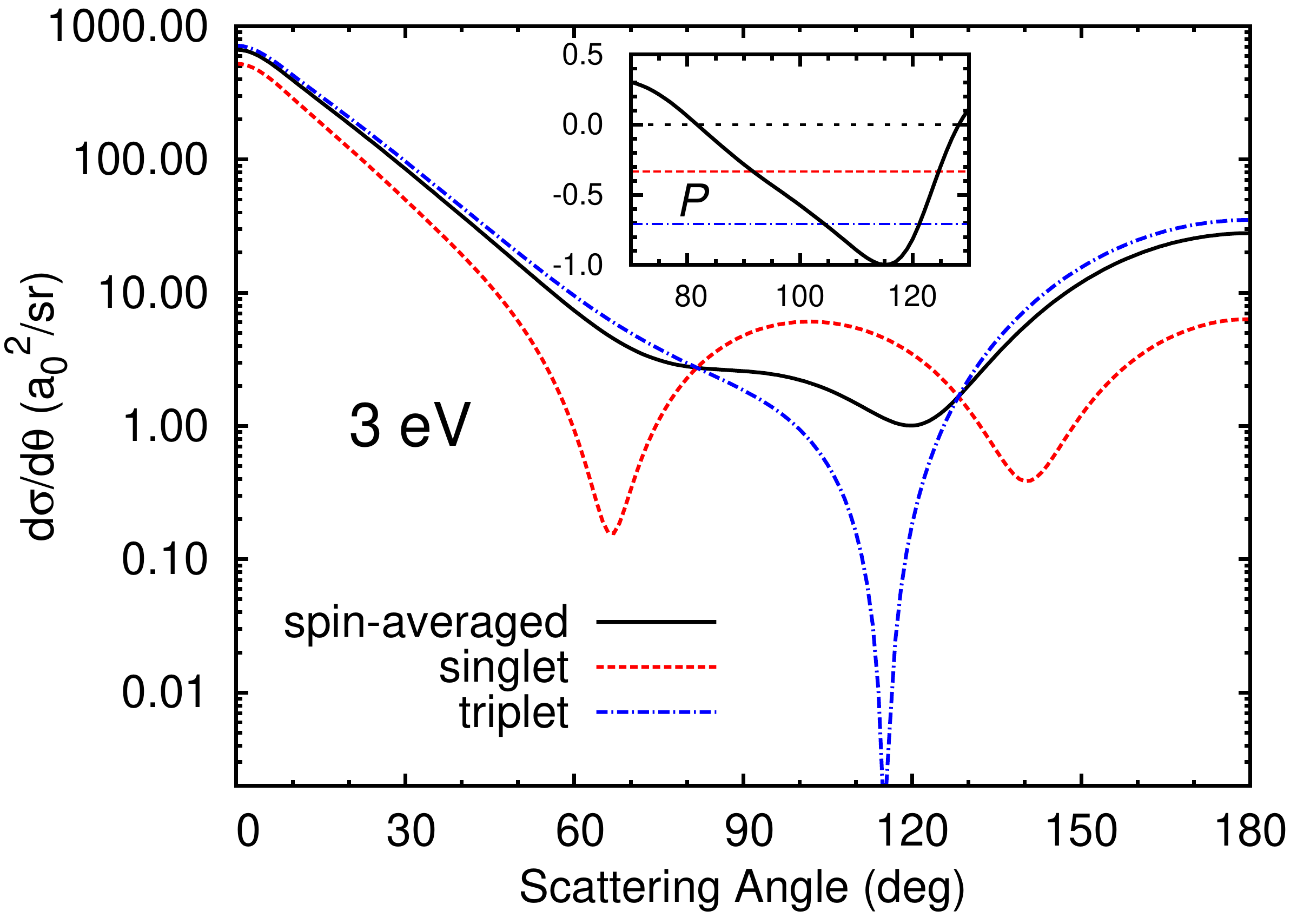}
\caption{Differential cross section as well as the individual contributions from the singlet and triplet total spin channels
for elastic electron scattering from Li atoms at a collision energy of 3~eV.
The insert shows the parameter~$P$ in the region $70^\circ-130^\circ$.  The lines at $-1/3$ and $-1/\sqrt2$ mark the
borders between separable and entangled as well as entangled and Bell-correlated regions, respectively.
}
\label{fig:3eV}
\end{figure}

Figure~\ref{fig:3eV} exhibits our results for a fixed energy of 3~eV as a function of the scattering angle.
Here one can see how the negative values of the spin correlation parameter develop.  Fortunately for a practical implementation, 
the DCS in the singlet spin channel assumes a local maximum around $110^\circ$, while the DCS for triplet
scattering assumes a deep minimum.  This explains why Baum {\it et al.}~\cite{Baum-AexLi} were able to
carry out measurements with small error bars in this angular regime.  

\begin{figure*}
\includegraphics[width=0.88\columnwidth]{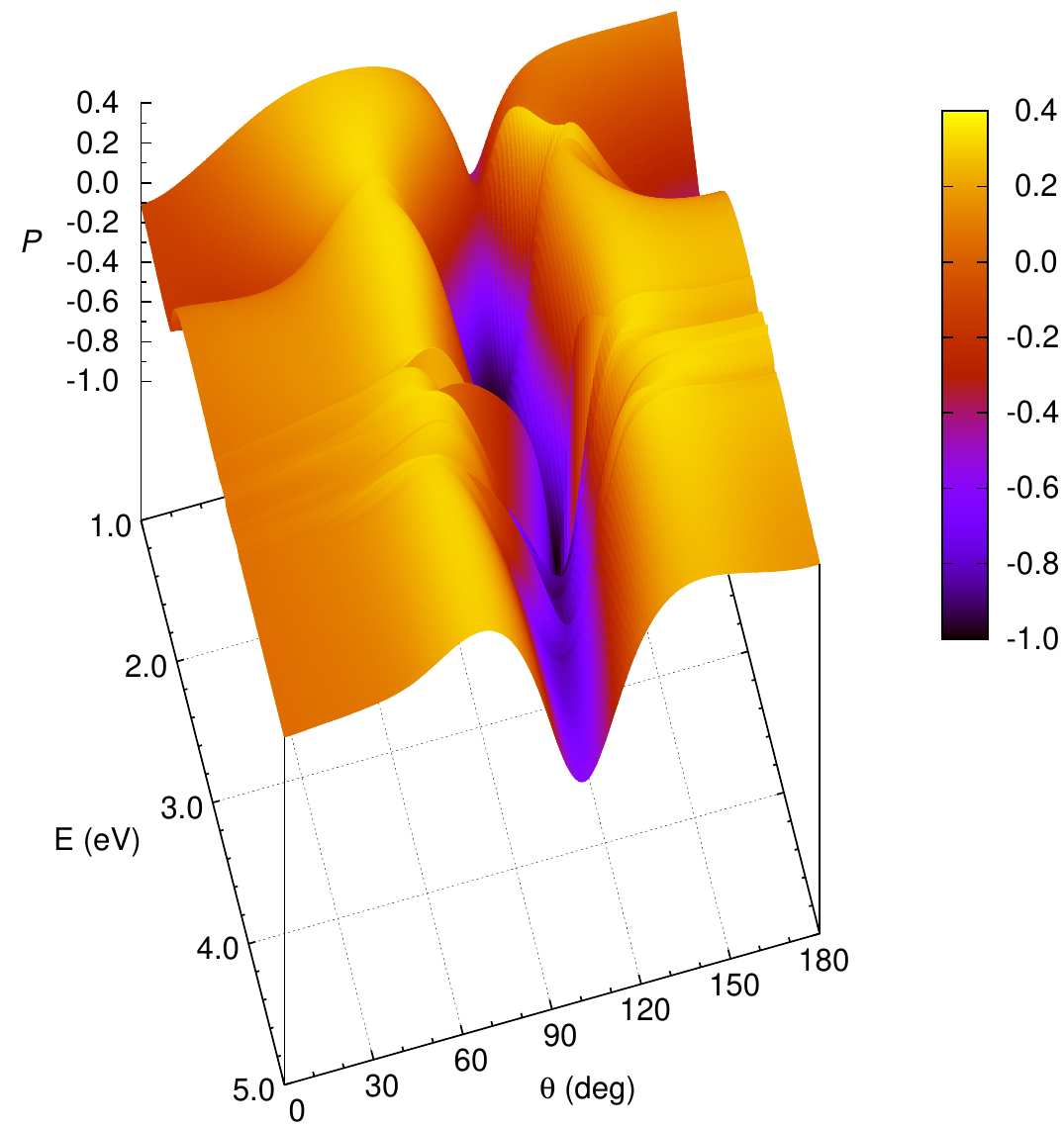}
~~~\includegraphics[width=1.08\columnwidth]{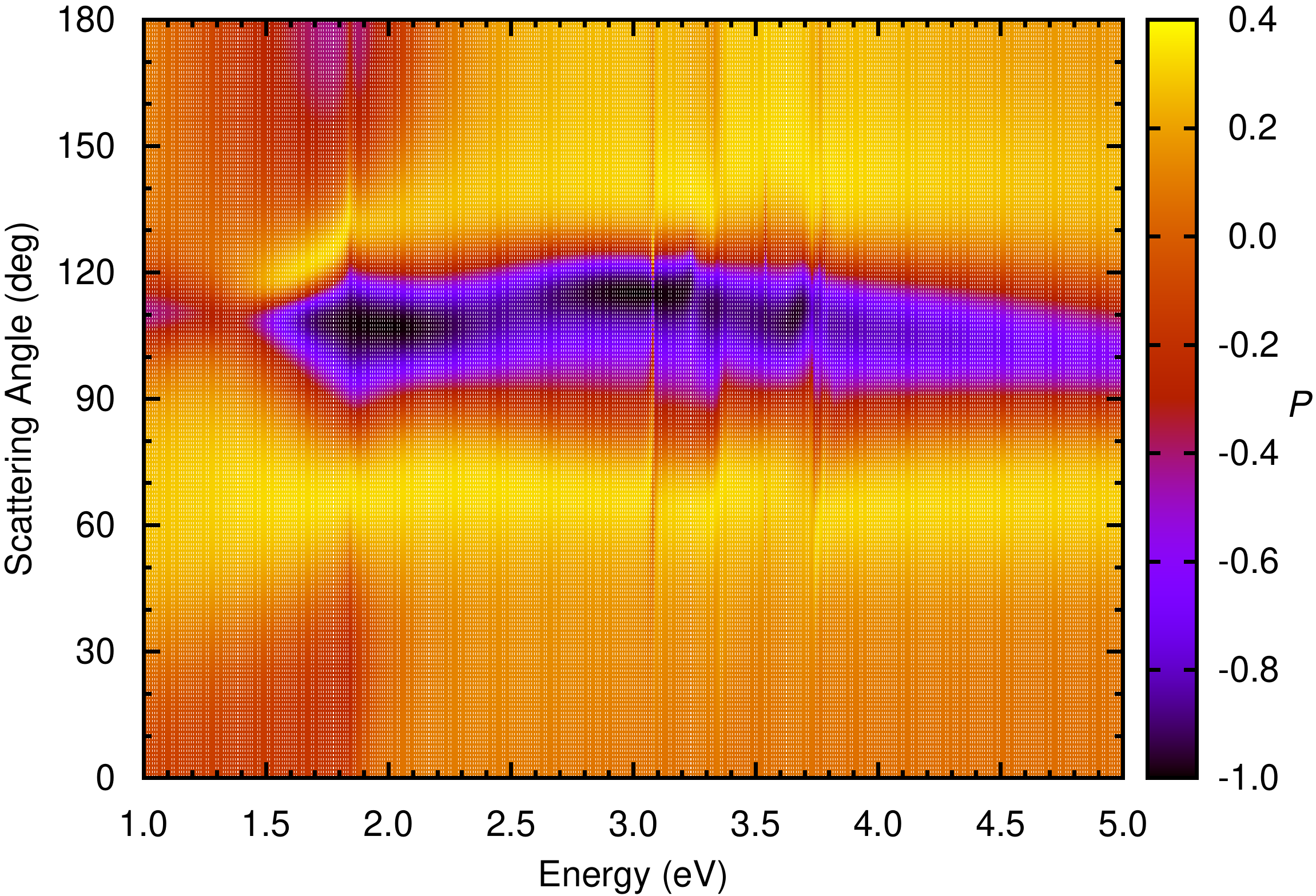}
\caption{Spin correlation parameter~$P$ for elastic electron scattering from lithium atoms, as 3D figure (left) and as contour plot (right). }
\label{fig:P}
\end{figure*}

\begin{figure*}
\includegraphics[width=0.88\columnwidth]{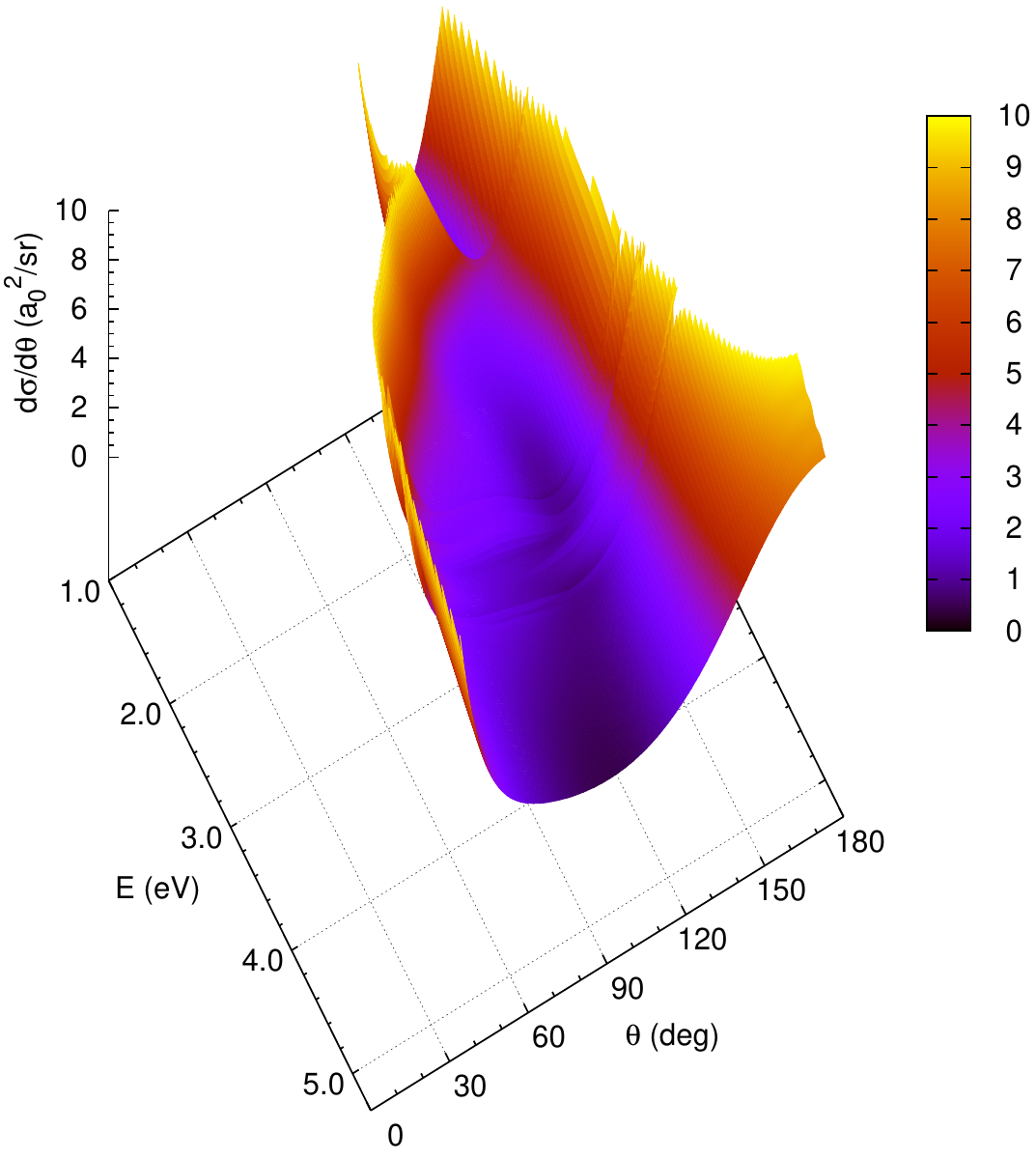}
~~~\includegraphics[width=1.08\columnwidth]{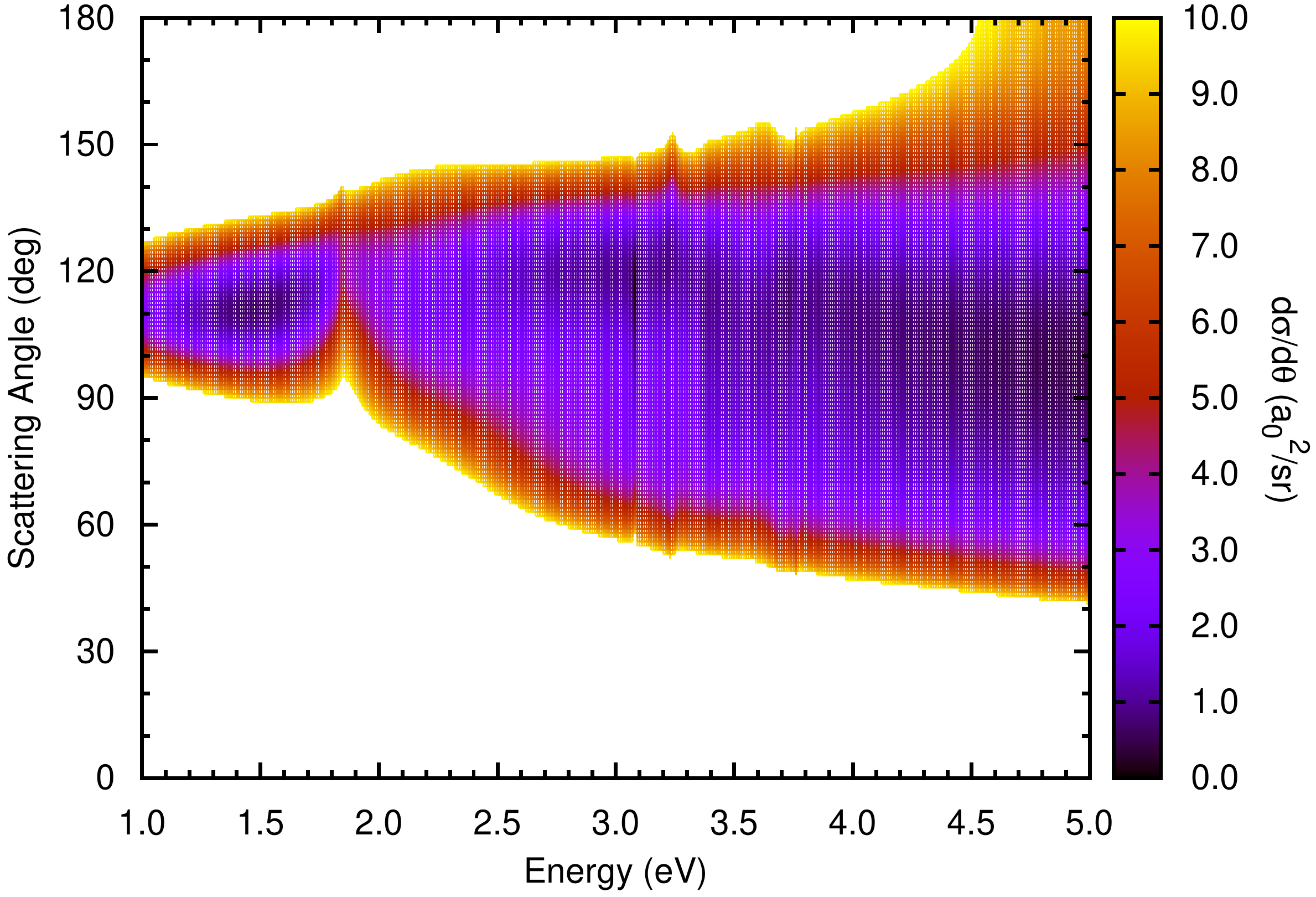}
\caption{Differential cross section for elastic electron scattering from lithium atoms, as 3D figure (left) and as contour plot (right).
         In the white areas of the contour plot, the DCS is larger than $10\,a_0^2$/sr. }
\label{fig:sigma}
\end{figure*}

As mentioned above, we are now in a position to provide a comprehensive overview of the results that might be expected
for the electron$-$lithium collision system.  This is done in Fig.~\ref{fig:P} for the spin correlation parameter and
Fig.~\ref{fig:sigma} for the DCS.  In the latter, we limit the maximum DCS value in the plots to $10\,a_0^2$/sr in order to
improve the visibility.  There is virtually no chance to find~$P$-values in the Bell-correlated regime when the DCS is
too large.   

To summarize: We have carried out calculations for elastic electron scattering from lithium atoms in a simple, but 
sufficient model to accurately predict the spin correlation parameter and the angle-differential cross section.
The most promising regime for Bell-correlations in this particular
collision system are energies between about 1.5~eV and 3.0~eV, in an angular range around $110^\circ \pm 10^\circ$.
While the cross sections are relatively small, the signal rate seems to be sufficient for a successful experimental implementation
of the scheme.
For higher energies than 3~eV, the results would first be affected by resonances.  Subsequently, as for
all scattering angles outside of the above range, triplet scattering is the dominant channel everywhere and hence \hbox{$P$-values} in
the Bell-correlated regime will not be achievable.  In the future, we plan to carry out similar calculations for 
atomic hydrogen and other alkali targets.

\medskip

One of us (K.B.) would like to thank Prof.\ K.~Blum and Dr.~B.~Lohmann for stimulating discussions.
This work was supported, in part, by the United States National Science Foundation under grant No.~PHY-1403245.

\end{document}